# Linear and Nonlinear Absolute Phase Effects in Interactions of Ulrashort Laser Pulses with a Metal Nano-Layer or with a Thin Plasma Layer

Author: **Sándor Varró**

Affiliation: Research Institute for Solid State Physics and Optics of the Hung. Acad. Sci.
Letters : H-1525 Budapest, POBox 49, Hungary
Phone : +36-1-392-2635,  Fax : +36-1-392-2215
E-mail : varro@sunserv.kfki.hu
Short title: Linear and Nonlinear Absolute Phase Effects
Number of pages: 18
Number of figures: 6

**Linear and Nonlinear Absolute Phase Effects in Interactions of Ulrashort Laser Pulses with a Metal Nano-Layer or with a Thin Plasma Layer**

Author: **Sándor Varró**

**Abstract.** It has been shown that in the scattered radiation, generated by an ultrashort laser pulse impinging on a metal nano-layer, non-oscillatory wake-fields appears with a definite sign. The magnitude of these wake-fields is proportional with the incoming field strength, and the definite sign of them is governed by the cosine of the carrier-envelope phase difference of the incoming pulse. When we let such a wake-field excite the electrons of a secondary target (say an electron beam, a metal surface or a gas jet), we can obtain 100 percent modulation in the electron signal in a given direction. This scheeme can serve as a basis for the construction of a robust linear carrier-envelope phase difference meter. At relativistic laser intensities the target is considered as a plasma layer in vacuum produced from a thin foil by a pre-pulse, which is followed by the main high-intensity laser pulse. The nonlinearities stemming from the relativistic kinematics lead to the appearance of higher-order harmonics in the scattered spectra. In general, the harmonic peaks are down-shifted due to the presence of an intensity-dependent factor. This phenomenon is analogous to the famous intensity-dependent frequency shift in the nonlinear Thomson scattering on a single electron. In our analysis an attention has also been paid to the role of the carrier-envelope phase difference of the incoming few-cycle laser pulse. It is also shown that the spectrum has a long tail where the heights of the peaks vary practically within one order of magnitude forming a quasi-continuum. By Fourier synthesising the components from this plateau region, attosecond pulses have been obtained.

**Key words:** few-cycle laser pulses, carrier-envelope phase difference, high-harmonic generation, relativistic laser-plasma interactions.

**PACS:** 41.20 Jb, 42.25 Bs, 52.38-r

# 1. Introduction

The study of the interaction of intense few-cycle laser pulses with matter has brought a new, important branch of investigations in nonlinear optics, as Brabec and Krausz (2000) emphasized in their review paper. The effect of the absolute phase (the carrier-envelope phase difference, in short: "absolute phase" or "CE phase") on the nonlinear response of atoms and of solids interacting with a very short, few-cycle strong laser pulse has recently drawn considerable attention and has initiated a wide-spreading theoretical and experimental research. For instance Paulus et al. (2001) have detected an anticorrelation in the shot-to-shot analysis of the photoelectron yield of ionization measured by two opposing detectors. This effect comes from the random variation of the CE phase (hence the direction and the magnitude of the electric field of the laser) from one pulse to the other. Such extreme short pulses can be used to monitor the details of photoelectron dynamics (Hentschel et al., 2001) or atomic inner-shell relaxation processes, like the Auger effect (Drescher et al., 2002). Concerning theory, the CE pase-dependence of the spatial asymmetry in photoionization has been investigated by Chelkowski and Bandrauk (2002), and by Milošević et al. (2002) (see also Milošević et al., 2003). In the meantime the problem of the stabilization and control of the CE phase in the few-cycle laser pulse trains has been achieved by Baltuška et al. (2003) and by Witte et al. (2004). On the basis of a simulation, using the time-dependent density functional approach, Lemell et al. (2003) predicted the CE phase-dependence of the photoelectron yield in case of the surface photoelectric effect of metals in the optical tunnel regime. Apolonskiy et al. (2004) and Dombi et. al. (2004) have reported the measurement of this effect, but the absolut phase-dependence had a considerably smaller modulation (CE phase-sensitivity) in their experiment than predicted by the simulation. Fortier et al. (2004) has recently demonstrated the CE phase effect in quantum interference of injected photocurrents in semiconductors. In the multiphoton regime Nakajima and Watanabe (2006) has found theoretically CE phase effects in the bound state population of a Cs atom excited by nearly single-cycle pulses.

    At this point we would like to note that there is a wide-spread opinion among researchers investigating the CE phase effects that these effects appear exclusively in *nonlinear* processes. In fact, as Fearn and Lamb, Jr. (1991) have shown already in 1991, the sine or the cosine character of the laser pulse makes a difference in the *linear* photoionization dynamics, if one takes into account the counter-rotating term in the interaction. As they wrote in Section IV. of their paper: "This suggests that…the time delay [of the electron signal] could be used to measure the phase of the field." A simple illustration of the linear CE phase effect has recently been considered by Ristow (2004) in the case of a harmonic oscillator.

    At relativistic incoming laser intensities the high-harmonic production on plasma surfaces has been considered analytically by Lichters et al. (1996) on the basis of their famous oscillating mirror model. The model introduced in the present paper considerably differs from this approach, because we have exactly taken into account the radiation reaction on the plasma layer. We note that recently there has been much labour put into the classical simulations of various processes (generation of coherent x-rays, laser acceleration of electrons) in laser-plasma interaction (see e.g. Pukhov & Meyer-ter-Vehn, 2003, Kiselev et al., 2004 and Quèrè et al., 2006). Moreover, the first experimental results by Hidding et al. (2006) appeared on the generation of quasi-monoenergetic electron bunches by strong laser fields. On the other hand, to our knowledge, there have been no classical relativistic considerations so far published, where the CE phase effect in the relativistic regime would have been analysed. The present paper may be considered as a contribution to the study of this particular aspect, too.

We have seen above, that in the theoretical works exclusively nonlinear quantum processes (photoionization, surface photoelectric effect) have been considered in the nonrelativistic regime. In the present paper we describe the reflection and transmission of a few-cycle laser pulse on a thin metal layer and a plasma layer represented by a surface current density of free electrons. Our analysis here, as in our earlier study (Varró, 2004), is based completely on classical electrodynamics and mechanics, in the frame of which we solve the system of coupled Maxwell-Lorentz equations of the incoming and scattered radiation and the surface current representing the metallic or plasma electrons.

In Section 2 we present the basic equations describing our model, and present the exact analytic solution of the scattering problem in the nonrelativistic regime. Here we shall briefly analyse the exact solutions in the frequency domain. We shall discuss the temporal behaviour of the reflected signal, and show that a pulse-decompression and "freezing-in" of the radiation field may happen, yielding to the appearance of a quasi-static wake-field in the scattered signal. In Section 3 we shall derive the relativistic equation of motion for the surface current density of the electrons which are considered the active charges in a thin plasma layer. It will turn out that the complete solution of the scattering problem can be reduced to the solution of a first order ordinary inhomogeneous differential equation. An approximate analytic solution to this equation will be given, which is valid for moderately relativistic incoming laser intensities. On the basis of these solutions the spectrum of the reflected radiation field containing higher-harmonics will be calculated. It will be shown that these spectra are considerably depend on the carrier-envelope phase difference of the incoming few-cycle laser pulse. In Section 4 a brief summary closes our paper.

## 2. The appearance of wake-fields in the scattered signal in the non-relativistic regime

The idea to study system under discussion appeared to us by reading a paper by Sommerfeld (1915) in which he analysed the temporal distortion of x-ray pulses of arbitrary shape and duration impinging perpendicularly on a surface current being in vacuum. In our earlier work (Varró, 2004) we have generalized this model in the following sense. On one hand, we allowed oblique incidence of the incoming radiation field, and on the other hand, we assume that the surface current distribution (which represents a thin metal layer) is embedded between two semi-infinite dielectrics with two different indeces of refraction. In the present paper we shall give the relativistic generalization of the equation of motion of the surface current, too, which was not investigated by Sommerfeld, either. This latter approach will be derived in Section 3, where we shall consider the interaction of a plasma layer with laser pulses of relativistic intensities. In the present Section we shall consider only the non-relativistic dynamics of the surface current.

### 2.1 The basic equations of the model and the exact analytic solutions in the frequency domain

The model to be used here we have already studied in our earlier work (Varró, 2004). For completeness of the present paper, let us first briefly summarize the basic notations and equations, which can also be found in this reference. We take the coordinate system such that the first dielectric with index of refraction $n_1$ fills the region $z > l_2 / 2$, this is called region 1. In region 2 we place the thin metal layer of thickness $l_2$ perpendicular to the z-axis and defined by the relation $-l_2 / 2 < z < +l_2 / 2$. Region 3, $z < -l_2 / 2$, is assumed to be filled by the second dielectric having the index of refraction $n_3$. The thickness $l_2$ is assumed to be much smaller then the average skin depth of the incoming radiation. The target defined this way can be imagined as a thin metal layer evaporated, for instance, on a glass substrate. This layer, in fact is assumed to be represented by a sheet of electrons bound to region 2 and moving freely in the x-y plane. In case of perpendicular incidence the light would come from the positive z-direction, and it would be transmitted in the negative z-direction into region 3. The plane of incidence is defined as the y-z plane and the initial $\vec{k}$-vector is assumed to make an angle $\theta_1$ with the z-axis. In case of an s-polarized incoming TE wave the components of the electric field and the magnetic induction read $(E_x,0,0)$ and $(0,B_y,B_z)$, respectively. They satisfy the Maxwell equations

$$\partial_y B_z - \partial_z B_y = \partial_0 \varepsilon E_x, \quad \partial_z E_x = -\partial_0 B_y, \quad -\partial_y E_x = -\partial_0 B_z \quad , \tag{1}$$

where $\varepsilon = n^2$ is the dielectric constant and $n$ is the index of refraction. If we make the replacements $\varepsilon E_x \to -B_x$, $B_z \to E_z$ and $B_y \to E_y$ then we have the field components of a p-polarized TM wave $(0,E_y,E_z)$ and $(B_x,0,0)$, and we receive the following equations

$$\partial_z B_x = \partial_0 \varepsilon E_y, \quad -\partial_y B_x = \partial_0 \varepsilon E_z, \quad \partial_y E_z - \partial_z E_y = -\partial_0 B_x \quad . \tag{2}$$

In the followings we will consider only the latter case, namely the scattering of a p-polarized TM radiation field. From Eq. (2) we deduce that $B_x$ satisfies the wave equation, and in region 1 we take it as a superposition of the incoming plane wave pulse $F$ and an unknown reflected plane wave $f_1$

$$B_{x1} = F - f_1 = F[t - n_1(y\sin\theta_1 - z\cos\theta_1)/c] - f_1[t - n_1(y\sin\theta_1 + z\cos\theta_1)/c]. \tag{3}$$

From Eq.(2) we can express the components $E_y$ and $E_z$ of the electric field strength by taking into account Eq. (3)

$$E_{y1} = (\cos\theta_1/n_1)(F + f_1), \quad E_{z1} = (\sin\theta_1/n_1)(F - f_1) \quad . \tag{4}$$

In region 3 the general form of the magnetic induction $B_{x3}$ is the by now unknown refracted wave $g_3$

$$B_{x3} = g_3 = g_3[t - n_3(y\sin\theta_3 - z\cos\theta_3)/c]. \tag{5}$$

The corresponding components of the electric field strength are expressed from the above equation with the help of the first two equation of Eq. (2)

$$E_{y3} = (\cos\theta_3/n_3)g_3, \quad E_{z3} = (\sin\theta_3/n_3)g_3. \tag{6}$$

In region 2 the relevant Maxwell equations with the current density $\vec{j}$ read

$$\partial_z B_x = (4\pi/c)j_{y2} + \partial_0 \varepsilon E_y, \quad \partial_y E_z - \partial_z E_y = -\partial_0 B_x. \tag{7}$$

By integrating the two equations in Eq. (7) with respect to z from $-l_2/2$ to $+l_2/2$ and taking the limit $l_2 \to 0$, we obtain the boundary conditions for the field components

$$[B_{x1} - B_{x3}]_{z=0} = (4\pi/c)K_{y2}, \quad [E_{y1} - E_{y3}]_{z=0} = 0, \tag{8}$$

where $K_{y2}$ is the y-component of the surface current in region 2. This surface current can be expressed in terms of the local velocity of the electrons in the metal layer

$$K_{y2} = e(d\delta_y/dt)l_2 n_e, \quad (4\pi/2c)K_{y2} = (m/e)\Gamma(d\delta_y/dt), \tag{9}$$

$$\Gamma \equiv 2\pi(e^2/mc)l_2 n_e, \quad \Gamma = (\omega_p/\omega_0)^2(\pi l_2/\lambda_0)\omega_0, \quad \kappa \equiv \Gamma/\omega_0 = \pi(\omega_p/\omega_0)^2(l_2/\lambda_0) \quad , \tag{10}$$

where for later convenience we have introduced $\omega_0$, $\lambda_0 = 2\pi c/\omega_0$, the carrier frequency and the central wavelength of the incoming light pulse, and $n_e, \omega_p = \sqrt{4\pi n_e e^2/m}$ denote the density of electrons and the corresponding plasma frequency in the metal layer, respectively. In Eq. (9) $\delta_y$ denotes the local displacement of the electrons in the metal layer for which we later write down the Lorentz equation (Newton equation in the non-relativistic regime) in the presence of the complete electric field. We remark that in reality the thickness $l_2$ is, of course, not infinitesimally small, rather, it has a finite value which is anyway assumed to be smaller then the skin depth $\delta_{skin} = c/\sqrt{\omega_p^2 - \omega_0^2}$. In order to have a feeling on the size of the parameters coming into our analysis, let us take some illustrative examples. For instance, for $n_e = 10^{22}/cm^3$, $\delta_{skin} = \lambda_0/14$, and for $n_e = 10^{23}/cm^3$, $\delta_{skin} = \lambda_0/47$, where we have taken $\lambda_0 = 800nm$ and $\omega_0 = 2.36 \times 10^{15} s^{-1}$ for a Ti:Sa laser. For the damping parameter $\Gamma$ in the first case, if we take $l_2 = \lambda_0/400 = 2nm$, we have $\kappa \equiv \Gamma/\omega_0 = 1/20$. In the second case, for the same thickness $l_2 = \lambda_0/400 = 2nm$ we have $\kappa \equiv \Gamma/\omega_0 = 0.18$.

From Eq. (8) with the help of Eq. (9) we can express $f_1$ and $g_3$ in terms of $\delta_y'(t')$

$$f_1(t') = (1/(c_1 + c_3))[(c_3 - c_1)F(t') - 2c_3(m/e)\Gamma \delta_y'(t')], \tag{11}$$

$$g_3(t') = (2c_1/(c_1 + c_3))[F(t') - (m/e)\Gamma \delta_y'(t')], \tag{12}$$

where the prime on $\delta_y$ denotes the derivative with respect to the retarded time $t' = t - yn_1 \sin\theta_1/c$ which is equal to $t - yn_3 \sin\theta_3/c$, securing Snell's law of refraction $n_1 \sin\theta_1 = n_3 \sin\theta_3$ to hold. Moreover, in Eqs. (11) and (12) we have introduced the notations $c_1 = \cos\theta_1/n_1$, $c_3 = \cos\theta_3/n_3$. We would like to emphasize that Eqs. (11) and (12) are valid in complete generality, that is, they hold for both non-relativistic and relativistic kinematics of the local electron displacement $\delta_y(t')$. For an interaction with a TM wave this displacement is uniform (along lines of constant *x*-values) in the direction perpendicular to the plane of incidence (the *y-z* plane), so it does not depend on the *x*-coordinate. As the incoming wave impinges on the surface at region 2 its (plane) wave fronts sweep this surface creating a superluminar polarization wave, described by the local displacement $\delta_y(t')$ of the electrons. Because of the continuity of $E_y$, Eq. (8), in the Newton equation for the displacement of the electrons in the surface current we can use for instance the force term $E_{y1} = c_1(F + f_1)$ according to Eq. (4), and neglect the magnetic induction. By taking Eq. (11) also into account we have

$$\delta_y''(t') = b[(e/m)F(t') - \Gamma \delta_y'(t')], \tag{13}$$

$$b \equiv 2c_1 c_3/(c_1 + c_3), \quad c_1 \equiv \cos\theta_1/n_1, \quad c_3 \equiv \cos\theta_3/n_3 = (1/n_3^2)\sqrt{n_3^2 - n_1^2(1 - \cos^2\theta_1)}. \tag{13a}$$

In obtaining the last formula in Eq. (13a) we have used Snell's law of refraction mentioned already a couple of lines before. For definiteness, we impose the initial conditions on the electron displacement $\delta_y(-\infty) = 0$ and $\delta_y'(-\infty) = 0$. Owing to Eqs. (11) and (12) the solution of Eq. (13) gives at the same time the complete solution of the scattering problem. We see that the equation of motion for the local displacement $\delta_y(t')$ contains a damping term with a damping parameter $b\Gamma$ where $\Gamma$ has been defined in Eq. (10). This latter constant is proportional with the squared plasma frequency and the thickness of the electron layer. The appearance of the damping term is a manifestation of the radiation reaction coming formally from the boudary conditions in the present description. Since $\Gamma$ is proportional with the electron density, this effect is due to the collective response of the electrons to the action of the complete (not only the incoming) radiation field, which, on the other had reacts back to the electrons. In the present description it is not possible to divide into steps these "consecutive" effects, as in the usual treatments of the radiation back-reaction. Eq. (13) can be solved exactly for an arbitrary incoming field $F(t)$. By calculating the Fourier transforms of Eqs. (11), (12) and (13) we can give an exact solution of the scattering problem in the frquency domain, namely

$$\tilde{f}_1(\omega) = -\frac{\tilde{F}(\omega)}{b\Gamma - i\omega}\left[b\Gamma + \frac{c_3 - c_1}{c_3 + c_1}i\omega\right] = -\frac{\tilde{F}(\omega)}{b\kappa - iv}\left[b\kappa + \frac{c_3 - c_1}{c_3 + c_1}iv\right], \tag{15}$$

$$\tilde{g}_3(\omega) = -\frac{2c_1}{c_1+c_3}\frac{i\omega\tilde{F}(\omega)}{b\Gamma - i\omega} = -\frac{2c_1}{c_1+c_3}\frac{i\nu\tilde{F}(\omega)}{b\kappa - i\nu}, \tag{16}$$

where we are using the dimensionless quantities $\kappa$ and $\nu$ given by the definitions

$$\kappa \equiv \Gamma/\omega_0, \quad \nu \equiv \omega/\omega_0, \tag{17}$$

and the geometrical factor $b$ was defined in Eq. (13a). It can be proved that the Fourier components of the reflected and the transmitted fluxes (to be calculated from Eqs. (15) and (16)) satisfy the following sum rule

$$c_1|\tilde{f}_1(\omega)|^2 + c_3|\tilde{g}_3(\omega)|^2 = c_1|\tilde{F}(\omega)|^2. \tag{18}$$

By now we have not specified the explicit form of the incoming field. Let us assume that it is a Gaussian quasi-monochromatic field with a carrier frequency $\omega_0$ having the carrier-envelope phase difference (CE phase) $\varphi$,

$$F(t) = F_0 \exp(-t^2/2\tau^2)\cos(\omega_0 t + \varphi), \tag{19}$$

where $\tau = \tau_L/2$ with $\tau_L$ being the full temporal width of the pulse's intensity. The Fourier transform of the incoming pulse given by Eq. (19) reads

$$\tilde{F}(\omega) \equiv \int_{-\infty}^{\infty} dt F(t)e^{i\omega t} = F_0\tau(\pi/2)^{1/2}\exp[-\tau^2(\omega^2+\omega_0^2)/2]\cdot\left(e^{i\varphi}e^{-\omega\omega_0\tau^2} + e^{-i\varphi}e^{+\omega\omega_0\tau^2}\right), \tag{20a}$$

and its modulus squared is the following $\pi$-periodic function of the CE phase

$$|\tilde{F}(\omega)|^2 = 2\{F_0\tau(\pi/2)^{1/2}\exp[-\tau^2(\omega^2+\omega_0^2)/2]\}^2(\cosh 2\tau^2\omega\omega_0 + \cos 2\varphi). \tag{20b}$$

From Eq. (20b) it is immediately seen that for any linear excitation process (whose response function is proportional with the incoming intensity) the frequency dependence of the modulation function ("visibility" or "contrast function") of the response function is given by the expression

$$M(\omega) = \frac{I_{max}(\omega) - I_{min}(\omega)}{I_{max}(\omega) + I_{min}(\omega)} = \frac{|\tilde{F}_{max}(\omega)|^2 - |\tilde{F}_{min}(\omega)|^2}{|\tilde{F}_{max}(\omega)|^2 + |\tilde{F}_{min}(\omega)|^2} = \frac{1}{\cosh 2\tau^2\omega\omega_0}, \tag{21}$$

where $I_{max,\,min}(\omega)$ here may mean both the maximum (minimum) values of the reflected flux $c_1|\tilde{f}_1(\omega)|^2$ and the refracted flux $c_3|\tilde{g}_3(\omega)|^2$ at a particular frequency $\omega$, as we vary the CE phase. This means that as we vary the CE phase between 0 and $\pi$, the modulation depth is given by Eq. (21). It is clear that if $\omega_0\tau$ is very large (which is the case of many-cycle pulses), then the modulation function is practically zero. On the other hand, when $\omega_0\tau$ is not large (which is the case of few-cycle pulses), then for small frequencies ($\omega\tau \ll 1$) we can have a modulation close to 100%. From Eqs. (15), (16) and (21) one obtains that the modulation of both the reflected and the transmitted signal is given by $1/\cosh 2\tau^2\omega\omega_0$ at a particular frequency. On the other hand neither of the reflection coefficient nor the transmission coefficient depend on the CE phase in the linear regime. It is clear that if the response function contains a resonance at a small frequency ($\omega\tau \ll 1$), then there is a better chance to observe the mentioned linear CE-phase-dependence at that particular resonance frequency.

At this point let us note that Eqs. (15) and (16) are more general than the original relations, Eqs. (11), (12) and (13), written in the time domain, where the indeces of refractions $n_1$ and $n_3$ have been taken as mere constants. The optional dispersion can be taken into account in Eqs. (15) and (16) by putting *by hand* a frequency dependence into $n_1$ and $n_3$. In the present paper we are not dealing with this aspect of the problem. Our description is approximate to the real scattering process in an other respect, too, namely it relies on the plane-wave model of the incoming, reflected and the refracted field. This is a standard approximation which is used in the texbooks throughout. Equations (15) and (16) can be superimposed for an assembly of incoming plane waves of different propagation direction, and this description would be suitable to handle the scattering of a *beam* with a limited transverse extension. In the present paper we are not dealing with this aspect of the scattering problem, either. Thus, our results are suitable to describe the scattering of unfocused beams.

**2. 2 The appearance of frozen-in wake-fields in the scattered signal**
As one sees from Eqs. (15) and (16) both the reflected and the refracted Fourier components have a pole at $\omega = -ib\Gamma$, which means that in the time domain a functional dependence of the form $\propto \exp(-b\Gamma t)$ is expected with a decay time $T/2\pi b\kappa$, where $b$ and $\kappa$ were defined in Eqs. (13a) and (17), respectively, and $T = 2\pi/\omega_0$ is the central period of the incoming field. This can easily be shown by integrating the ordinary first order differential equation Eq. (13) for the local velocity $\delta'_y(t')$, which determines through Eq. (11) and (12) the scattered fields,

$$\delta'_y(t') = e^{-b\Gamma t'} \int_{-\infty}^{t'} du [b(e/m)F(u)] e^{+b\Gamma u} . \tag{22}$$

The integral in Eq. (22) can be easily evaluated for any usual functional form of the incoming field. From Eq. (11) and (22) the completete reflected field can be obtained,

$$f_1(t') = \frac{c_3 - c_1}{c_3 + c_1} F(t') - 2\pi b^2 \kappa \exp[-2\pi b\kappa(t'/T)] \int_{-\infty}^{t'} d(u/T) F(u) \exp[+2\pi b\kappa(u/T)], \tag{23a}$$

where now in region 1 the retarded time reads
$$t' = t - n_1(y \sin\theta_1 + z \cos\theta_1)/c . \tag{23b}$$

For a model pulse of a $\sin^2$ envelope of finite support (which has been widely used in numerical simulations), the integral can be calculated analytically. Henceforth, in our illustrative numerical examples we will use the Gaussian pulse defined in Eq. (19).

     Before entering into some numerical examples of the temporal behaviour of the scattered field, let us make some general remarks. It is clear that if the pulse duration of the incoming field $F(t)$ is much smaller than the characteristic time $1/b\Gamma = T/2\pi b\kappa$ of the exponential factor, then $F(t)$ cuts off a small region in the inegration interval in Eq. (22) because of the relative smoothness of the exponential function. This means that the upper limit of the integral can be extended up to infinity within a reasonable approximation. Hence the *asympthotic behaviour of the velocity* $\delta'_y(t')$ can certainly be well represented by the approximate formula

$$\delta'_y(t') \approx e^{-b\Gamma t'} \int_{-\infty}^{+\infty} du [b(e/m)F(u)] e^{+b\Gamma u} = e^{-b\Gamma t'} b(e/m) \widetilde{F}(-ib\Gamma) , \tag{24a}$$

where $\widetilde{F}(-ib\Gamma)$ denotes the Fourier transform of the incoming field. By taking Eq. (20a) into account, we obtain from Eq. (23a)

$$\delta'_y(t') \approx e^{-b\Gamma t'} b(e/m)\left(F_0 \tau\sqrt{2\pi}\right) \exp\left[-(\omega_0^2 \tau^2/2)(1 - b^2\kappa^2)\right] \cos(\varphi + \omega_0^2 \tau^2 b\kappa), \tag{24b}$$

$$\delta'_y(t')/c$$
$$\approx -\left(b\omega_0\tau\sqrt{2\pi}\right)\exp\left[-(\omega_0^2\tau^2/2)(1-b^2\kappa^2)\right] \cdot \mu \exp\left(-2\pi b\kappa \frac{t'}{T}\right) \cos(\varphi + \omega_0^2\tau^2 b\kappa), \tag{24c}$$

where we have introduced the usual intensity parameter $\mu$ with the definition

$$\mu \equiv \frac{eF_0}{mc\omega_0} = 10^{-9} \sqrt{I}/E_{ph}, \quad \mu^2 = 10^{-18} I\lambda^2 . \tag{24d}$$

In the numerical expressions in Eq. (24d) $I$ denotes the mean intensity of the pulse measured in W/cm$^2$, $E_{ph}$ is the mean photon energy in eV-s, and $\lambda$ denotes the central wavelength of the pulse measured in microns (we have not displayed the numerical prefactors of order unity). According to Eq. (24c), the polarization current and (due to Eq. (11)) the scattered field itself contains a *wake-field* which is a *frozen-in non-oscillatory quasi-static field* propagating in the direction $(0, \sin\theta_1, \cos\theta_1)$ after leaving the metal layer. The decay time of this wake-field is given by $T/2\pi b\kappa$, which can be much larger than an optical period, as is shown on **Fig. 1**, where we have taken $\kappa = 1/20$. Since we are dealing just with few-cycle incoming pulses, this means, that the wake-field is still present, after the main pulse has passed. Moreover, from Eq. (23a) it can be seen, that at the Brewster angle of incidence ($c_1 = c_3 \to \theta \approx 56°$ if $n_1 = 1$, $n_3 = 1.5$) only the wake-field is propagating in that particular direction. It is remarkable, that, as can be seen from Eqs. (24b-c), the amplitude of the wake-field is proportional with the cosine of the CE phase $\varphi$. Thus, for instance, by varying the CE phase, the sign of the quasi-static wake-field can be reversed. On **Fig. 2** an illustrative example is shown for the temporal behavior of the reflected signal at Brewster angle of incidence of an incoming one-cycle Ti:Sa laser pulse.

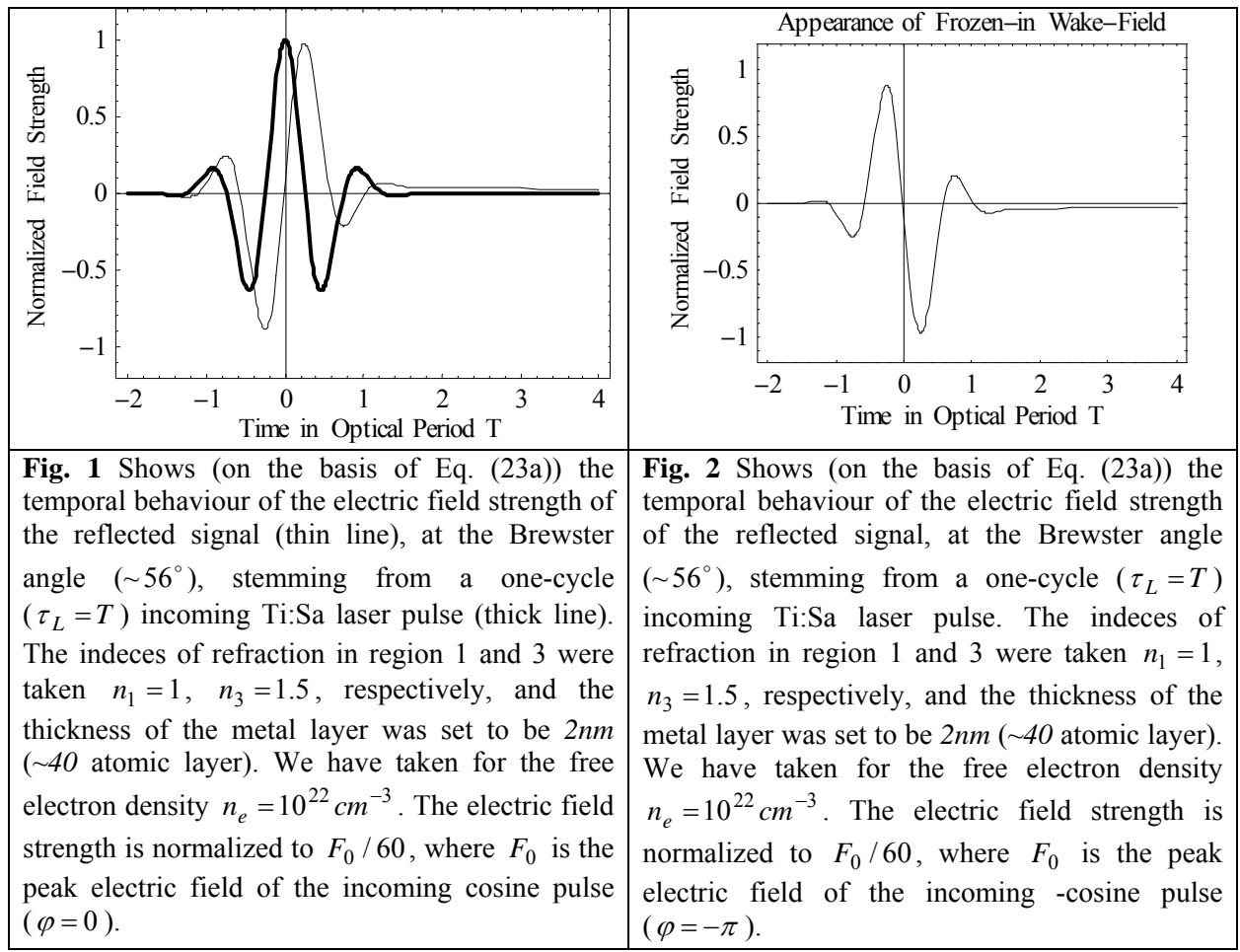

**Fig. 1** Shows (on the basis of Eq. (23a)) the temporal behaviour of the electric field strength of the reflected signal (thin line), at the Brewster angle ($\sim 56°$), stemming from a one-cycle ($\tau_L = T$) incoming Ti:Sa laser pulse (thick line). The indeces of refraction in region 1 and 3 were taken $n_1 = 1$, $n_3 = 1.5$, respectively, and the thickness of the metal layer was set to be *2nm* ($\sim 40$ atomic layer). We have taken for the free electron density $n_e = 10^{22} cm^{-3}$. The electric field strength is normalized to $F_0/60$, where $F_0$ is the peak electric field of the incoming cosine pulse ($\varphi = 0$).

**Fig. 2** Shows (on the basis of Eq. (23a)) the temporal behaviour of the electric field strength of the reflected signal, at the Brewster angle ($\sim 56°$), stemming from a one-cycle ($\tau_L = T$) incoming Ti:Sa laser pulse. The indeces of refraction in region 1 and 3 were taken $n_1 = 1$, $n_3 = 1.5$, respectively, and the thickness of the metal layer was set to be *2nm* ($\sim 40$ atomic layer). We have taken for the free electron density $n_e = 10^{22} cm^{-3}$. The electric field strength is normalized to $F_0/60$, where $F_0$ is the peak electric field of the incoming -cosine pulse ($\varphi = -\pi$).

On **Fig. 2** it is seen that the magnitude of the electric field of the wake-field is about 600 times smaller than the peak field strength of the incoming cosine pulse. The peak electric field strength $F_0$ of a laser field can be calculated from the laser intensity $I_0$ according to the following formula,

$$[F_0 /(V/cm)] = 27.46 \times [I_0 /(W/cm^2)]^{1/2} . \tag{25}$$

Thus, for an incoming laser of intensity $I_0 \approx 10^{12} W/cm^2$, the amplitude of the frozen-in wake-field in the reflected signal is of oder of 50000 *V/cm*, which is quite a large value for a quasi-static field. As we have already mentioned above the amplitude and sign of this wake-field can be varied by changing the CE phase of the incoming pulse. The idea naturally emerges, that when we let such a wake-field excite the electrons of a secondary target – say an electron beam, a second metal plate or a gas jet – *we may obtain 100 percent modulation depth in the electron signal (acceleration, tunnel-ionization) in a given direction, even for a relatively low intensity incoming field.* This scheeme can perhaps serve as a basis for the construction of a robust linear carrier-envelope phase difference meter. At the end of the present section we would like to emphasize that the generation of the wake-fields discussed above is a linear process, since the amplitude of the wake-field is simply proportional with the amplitude of the incoming field. This means that for the observation of the effect considered here one does not need extremely high laser intensities.

## 3. High-harmonic generation on a plasma layer in the relativistic regime

In the present Section we shall first derive the relativistic equation of motion of the surface current density of the electrons which are considered the active charges in a thin plasma layer. It will turn out that the complete solution of the scattering problem can be reduced to the solution of a first order ordinary inhomogeneous differential equation. We shall give an approximate analytic solution to this equation valid for moderately relativistic incoming laser intensities (where $\mu < 1$). On the basis of these solutions we shall calculate the spectrum of the reflected radiation field.

### 3.1 Relativistic equation of motion of the surface current density of the electrons

Henceforth, in the equation of motion for the surface current $K_{y2}$, Eq. (9), we use $B_{x3}$, $E_{y3}$ and $E_{z3}$, Eq. (5) and (6), and take into account Eq. (12). Moreover, we specialize our system to be a plasma layer in vacuum (hence $n_1 = n_3 = 1 \to \theta_1 = \theta_3 = \theta$). If we take into account retardation, i.e. relativistic effects, the argument of the field strength will be given in the layer as

$$\eta \equiv \left[ t - \frac{(y+\delta_y)\sin\theta - (z+\delta_z)\cos\theta}{c} \right]_{z=0} . \tag{26}$$

The local displacements $\delta_y$ and $\delta_z$ in the layer will depend on the retarded time $t' = t - y\sin\theta/c$. In fact, these displacements represent a polarization wave in the layer propagating in the positive *y*-direction with the velocity $c/\sin\theta$. This is due to the assumed oblique incidence of the incoming field, i.e. the wave front of the incoming field sweeps the layer with such a velocity. The fields to be used in the equation of motion of the electron displacements can be expressed as $\vec{E} = \vec{\varepsilon} g_3$ and $\vec{B} = g_3(1,0,0) = \vec{n} \times \vec{E}$, where $\vec{\varepsilon} = (0,\cos\theta,\sin\theta)$ is the polarization vector and $\vec{n} = (0,\sin\theta,-\cos\theta)$ is the unit propagation vector. The true retarded time parameter is given from Eq. (26) as $\eta = t' - \vec{n}\cdot\vec{\delta}/c$ where $\vec{\delta} = (0,\delta_y,\delta_z)$. Introducing the velocity $\vec{v} = d\vec{\delta}/dt'$ and the associated relativistic factor $\gamma = 1/\sqrt{1-v^2/c^2}$ we can define the proper time element $d\tau = dt'/\gamma$. Moreover we define the four-position $\delta^\mu = \{ct', \vec{\delta}\}$ and the four-velocity $u^\mu = d\delta^\mu/d\tau = \{u_0, \vec{u}\}$. In this way the following set of relativistic equations derives for the four-velocity associated to the electron displacements in the plasma layer

$$m(d\vec{u}/d\tau) = (e/c)(u_0 \vec{E} + \vec{u} \times \vec{B}) \text{, and } m(du_0/d\tau) = (e/c)\vec{u} \cdot \vec{E}. \tag{27}$$

We stress that in Eq.(15) the field strengths depend on the true retarded time parameter $\eta = t' - \vec{n} \cdot \vec{\delta}/c$ at the position of the electrons, where $t' = t - y\sin\theta/c$ is the uniform retarded time over the plasma layer. We note that, because of the assumed geometry of the scattering, the electrons move collectively in phase along the lines parallel to the *x*-axis. The second of the two equations in Eq. (27) expresses the relativistic work theorem. From the two equations in Eq. (27) there can be derived the important relations

$$u_0 - \vec{n} \cdot \vec{u} = ca = const. \,, \tag{28}$$

and, as a consequence, on the other hand

$$cd\eta/d\tau = \frac{d}{d\tau}[ct' - \vec{n} \cdot \vec{\delta}(t')] = u_0 - \vec{n} \cdot \vec{u} = ca, \quad d\eta/d\tau = a = \gamma(1 - \vec{n} \cdot \vec{\upsilon}/c), \tag{29}$$

where *a* is a constant depending on the initial local velocity. This means, that the derivatives with respect to the proper time are proportional to the derivatives with respect to the argument of the field strengths, $d/d\tau = ad/d\eta$, where the constant *a* depends only on the initial conditions. As is seen from the last relation in Eq. (29), for a particle initially at rest this constant is $a = 1$. According to the relation $d/d\tau = ad/d\eta$ just shown, the equation of motion in Eq. (27) can be brought to the form

$$\frac{d^2\vec{\delta}}{d\eta^2} = \frac{e}{ma}\left[\vec{E} + \vec{n}\left(\frac{1}{c}\frac{d\vec{\delta}}{d\eta} \cdot \vec{E}\right)\right], \tag{30}$$

where, as we saw before $\vec{E} = \vec{\varepsilon} g_3$, and, according to Eq. (12)

$$g_3 = F - (m/e)\Gamma(a/\gamma)(d\delta_y/d\eta). \tag{31}$$

Now let us make the following decomposition of the displacement $\vec{\delta}$,

$$\delta_\perp = \vec{\varepsilon} \cdot \vec{\delta} = \delta_y \cos\theta + \delta_z \sin\theta \,, \quad \delta_\parallel = \vec{n} \cdot \vec{\delta} = \delta_y \sin\theta - \delta_z \cos\theta \,, \tag{32}$$

that is

$$\delta_y = \delta_\perp \cos\theta + \delta_\parallel \sin\theta \,, \quad \delta_z = \delta_\perp \sin\theta - \delta_\parallel \cos\theta \,. \tag{33}$$

With the help of this decomposition and by integration with respect to $\eta$, it can be seen from Eq. (30) that

$$d\delta_\parallel/d\eta = \frac{1}{2c}[(d\delta_\perp/d\eta)^2 - (d\delta_\perp/d\eta)_0^2]. \tag{34}$$

Henceforth we will take the initial value $(d\delta_\perp/d\eta)_0 = 0$ which corresponds to $a = 1$. In this way we obtain

$$d\delta_y/d\eta = (d\delta_\perp/d\eta)\cos\theta + (1/2c)(d\delta_\perp/d\eta)^2 \sin\theta. \tag{35}$$

Hence, if we solve the equation of motion for $d\delta_\perp/d\eta$, we can express $d\delta_y/d\eta$ through which, according to Eq. (31), the transmitted field can be calculated. Similarily, because of Eq.(11), the reflected field can also be determined.

Combining Eqs. (31) and (23) – after some lengthy but straightforward algebra – we receive the following closed equation for $\delta_\perp$

$$\frac{d^2\delta_\perp}{d\eta^2} = \frac{e}{m}F(\eta) - \Gamma \frac{(d\delta_\perp/d\eta)\cos\theta + [(d\delta_\perp/d\eta)^2/2c]\sin\theta}{\sqrt{1 + (d\delta_\perp/d\eta)^2/c^2 + (d\delta_\perp/d\eta)^4/4c^4}}. \tag{36}$$

In case of perpendicular incidence ($\theta = 0$), according to Eq. (21), we directly obtain an equation for $\delta_y$, with which we have to express the scattered fields,

$$\frac{d^2\delta_y}{d\eta^2} = \frac{e}{m} F(\eta) - \Gamma \frac{(d\delta_y/d\eta)}{\sqrt{1+(d\delta_y/d\eta)^2/c^2 + (d\delta_y/d\eta)^4/4c^4}}. \tag{37}$$

We note that both Eq. (36) and Eq. (37), in fact, are first-order differential equations for $d\delta_\perp/d\eta$.

**3. 1 High-harmonic generation on a plasma layer in the moderate relativistic regime**

By now, the form of the incoming pulse $F(\eta)$ has not been specified, it can be of arbitrary shape. As is seen from Eqs.(11) and (12), both the reflected and the transmitted signal contain the the unknown term $d\delta_y/dt'$. Due to Eqs. (35), (36) or (37), if once we know $d\delta_y/d\eta$, then the Fourier components of this unknown quantity $d\delta_y/dt'$ can be expressed as

$$\frac{d\delta_y}{dt'}(\omega) = \int_{-\infty}^{+\infty} d\eta \frac{d\delta_y}{d\eta} \exp\{i\omega[\eta + \delta_\parallel(\eta)/c]\}, \tag{38}$$

where we have taken into account the relation $\eta = t' - \vec{n}\cdot\vec{\delta}/c = t' - \delta_\parallel/c$. In this way the solution of the scattering problem is reduced to the solution of the (non-linear) ordinary differential equation Eq. (36) (or, in case of perpendicular incidence, Eq. (37)).

In the present section we consider moderately relativistic motions in the plasma layer, hence we approximate the square root by unity and neglect the second term in the nominator in Eq. (37). The resulting equation for $d\delta_\perp/d\eta$ yields

$$\frac{d^2\delta_\perp}{d\eta^2} = \frac{e}{m} F(\eta) - (\Gamma\cos\theta)\frac{d\delta_\perp}{d\eta}, \tag{39}$$

which is formally coincides with Newton equation, Eq. (13), suitable for the non-relativistic description.

Henceforth, for simplicity, we denote $\eta$ by $t$. Assuming an impinging Gaussian pulse

$$F(t) = F_0 \exp(-t^2/2\tau^2)\cos(\omega_0 t + \varphi) \tag{40}$$

of amplitude $F_0$, carrier frequency $\omega_0$, pulse width $\tau = \tau_L/2$ (where $\tau_L$ is the width of the intensity), and of carrier-envelope phase difference $\varphi$, Eq. (39) can be approximately solved, yielding

$$d\delta_\perp/dt \approx (eF_0/m\omega_0)\left(1/\sqrt{1+\Gamma^2\cos^2\theta/\omega_0^2}\right)\exp(-t^2/2\tau^2)\sin(\omega_0 t + \varphi + \alpha), \tag{41}$$

where the additional phase $\alpha$ is defined by the relation

$$\sin\alpha = \frac{(\Gamma/\omega_0)\cos\theta}{\sqrt{1+(\Gamma/\omega_0)^2\cos^2\theta}}. \tag{42}$$

The phase term $\delta_\parallel = \vec{n}\cdot\vec{\delta}$ appearing in the exponential in Eq. (38) can be calculated from Eq. (41) by using Eq. (34) to yield

$$\delta_\parallel/c \approx \frac{1}{2}\left(\frac{eF_0}{mc\omega_0}\right)^2 \frac{1}{1+(\Gamma/\omega_0)\cos^2\theta} \\ \times \left\{\frac{1}{2}\int_{-\infty}^{t} dx \exp(-x^2/\tau^2) - \frac{1}{4\omega_0}\exp(-t^2/\tau^2)\sin[2(\omega_0 t + \varphi + \alpha)]\right\}. \tag{43}$$

According to Eq. (35), on the basis of Eq. (41), $d\delta_y / dt$ can be approximately expressed as

$$d\delta_y / cdt \approx 2\beta \cos\theta \exp(-t^2/2\tau^2) \sin(\omega_0 t + \varphi + \alpha)$$
$$+ \beta^2 \sin\theta \exp(-t^2/\tau^2)\{1 - \cos[2(\omega_0 t + \varphi + \alpha)]\}$$
(44)

where we have introduced the dimensionless parameter

$$\beta \equiv \frac{1}{2}\left(\frac{eF_0}{mc\omega_0}\right) \frac{1}{\sqrt{1 + (\Gamma/\omega_0)^2 \cos^2\theta}} = \frac{\mu}{2} \frac{1}{\sqrt{1 + (\Gamma/\omega_0)^2 \cos^2\theta}},$$
(45)

where $\mu \equiv eF_0/mc\omega_0$ is the dimensionless intensity parameter (which has already been defined in Eq. (24b)) usually appearing in strong field phenomena.

The main problem in calculating the scattered (e. g. the reflected) spectrum is the presence of the time integral in Eq. (43), which is – according to Eq. (38) – is present in the exponential of the Fourier integral. In order to get rid of this difficulty, we approximate this time integral in the following manner

$$\int_{-\infty}^{t} dx \exp(-x^2/\tau^2) \approx \begin{cases} 0, & -\infty \leq t \leq -\tau\sqrt{\pi}/2 \\ \tau\sqrt{\pi}/2 + t, & -\tau\sqrt{\pi}/2 \leq t \leq \tau\sqrt{\pi}/2 \\ \tau\sqrt{\pi}, & \tau\sqrt{\pi}/2 \leq t \leq +\infty \end{cases}.$$
(46)

We have numerically checked that the right hand side of Eq.(46) quite reasonably approximates the integral on the left hand side. We also note that for $\tau \to \infty$ (which corresponds to a very long laser pulse), only the second range gives a contribution. Accordingly, we split the time integral in Eq. (38) into three parts

$$\frac{d\delta_y}{dt'}(\omega) = \int dt \frac{d\delta_y(t)}{dt} \exp\{i\omega[t + \delta_\parallel(t)/c]\} \approx$$
$$\int_{-\infty}^{-\tau\sqrt{\pi}/2} dt B(t) e^{i\omega t} + e^{i\beta^2 \tau\sqrt{\pi}\omega/2} \int_{-\tau\sqrt{\pi}/2}^{+\tau\sqrt{\pi}/2} dt B(t) e^{i(1+\beta^2)\omega t} + e^{i\beta^2 \tau\sqrt{\pi}\omega} \int_{+\tau\sqrt{\pi}/2}^{+\infty} dt B(t) e^{i\omega t},$$
(47)

where

$$B(t) \approx \sum_n J_n(\beta^2 \omega / 2\omega_0)\{2\beta \cos\theta [e^{-i(2n-1)(\varphi+\alpha)} e^{-(n+1/2)t^2/\tau^2} e^{-i(2n-1)\omega_0 t} -$$
$$e^{-i(2n+1)(\varphi+\alpha)} e^{-(n+1/2)t^2/\tau^2} e^{-i(2n+1)\omega_0 t}]/2i + \beta^2 \sin\theta[2e^{-i2n(\varphi+\alpha)} e^{-(n+1)t^2/\tau^2} e^{-i2n\omega_0 t} -$$
$$e^{-i(2n-2)(\varphi+\alpha)} e^{-(n+1)t^2/\tau^2} e^{-i(2n-2)\omega_0 t} - e^{-i(2n+2)(\varphi+\alpha)} e^{-(n+1)t^2/\tau^2} e^{-i(2n+2)\omega_0 t}]/2\}$$
(48)

In obtaining Eq. (48) we have used the Jacobi-Anger formula, the generating formula for the Bessel functions, $e^{-iz\sin\psi} = \sum_n J_n(z) e^{-in\psi}$, moreover we have made the approximation $J_n[(\beta^2\omega/2\omega_0)e^{-t^2/\tau^2}] \approx J_n(\beta^2\omega/2\omega_0)e^{-nt^2/\tau^2}$, which means that – concerning the time dependence – we keep only the leading term in the power expansion of the Bessel functions $J_n$. This is a reasonable approximation at moderately relativistic intensities ($\mu < 1$). Within these approximations the integrals in Eq. (47) can be calculated by using the formulae 3.322.1-2 of Gradshteyn and Ryzhik (1980). We shall not list these lengthy formulae in the present paper.

We have checked that for relatively long pulses, say, for a 10-cycle pulse, at a relativistic intensity $2\times 10^{19} W/cm^2$, only the second term in Eq. (47) contributes considerably to the spectrum whose maxima correspond to the *intensity-dependend frequency-shifted harmonics* of frequencies $\omega_n = n\omega_0 /(1+\beta^2)$, where $\beta$ was defined in Eq. (45). We have found that the spectrum has a very *sharp cut-off determined by the critical index* depending here on the factor $\beta^2/2$ in the argument of the Bessel function. We can borrow a formula for this critical index $n_c$ from the theory of synchrotron radiation (see e.g. Jackson, 1962); $n_c = 3/(1-\beta^4/4)^{3/2}$, whose numerical value is approximately 78 in the case of 45° of incidence. The critical normalized frequency becomes $v_c = \omega_c/\omega_0 = n_c/(1+\beta^2) \approx 28$ which is in agreement whith what we have seen from our numerical calculations. Of course, for this estimate we have to assume that $\beta^2/2 < 1$, namely that the mentioned factor in the argument of the Bessel function is close to, but smaller than one.

To show the spectra for short, 2-cycle pulses we assume the electron density $n_e = 10^{21} cm^{-3}$ and thickness $l_2 = \lambda/100 = 8nm$ for the plasma layer as above, but we take a "moderate" intensity, namely $2 \times 10^{18} W/cm^2$, so one order of magnitude smaller as in the previous example. Then we obtain $v_n = (\omega_n/\omega_0) = n/(1+\beta^2) = \{0.84,\ 1.68,\ 2.51,\ 3.35\}$ for the first four harmonics $n = 1,\ 2,\ 3,\ 4$, where the parameter $\beta$ defined in Eq. (45) is proportional with the usual intensity parameter $\mu = 10^{-9} I^{1/2} \lambda$.

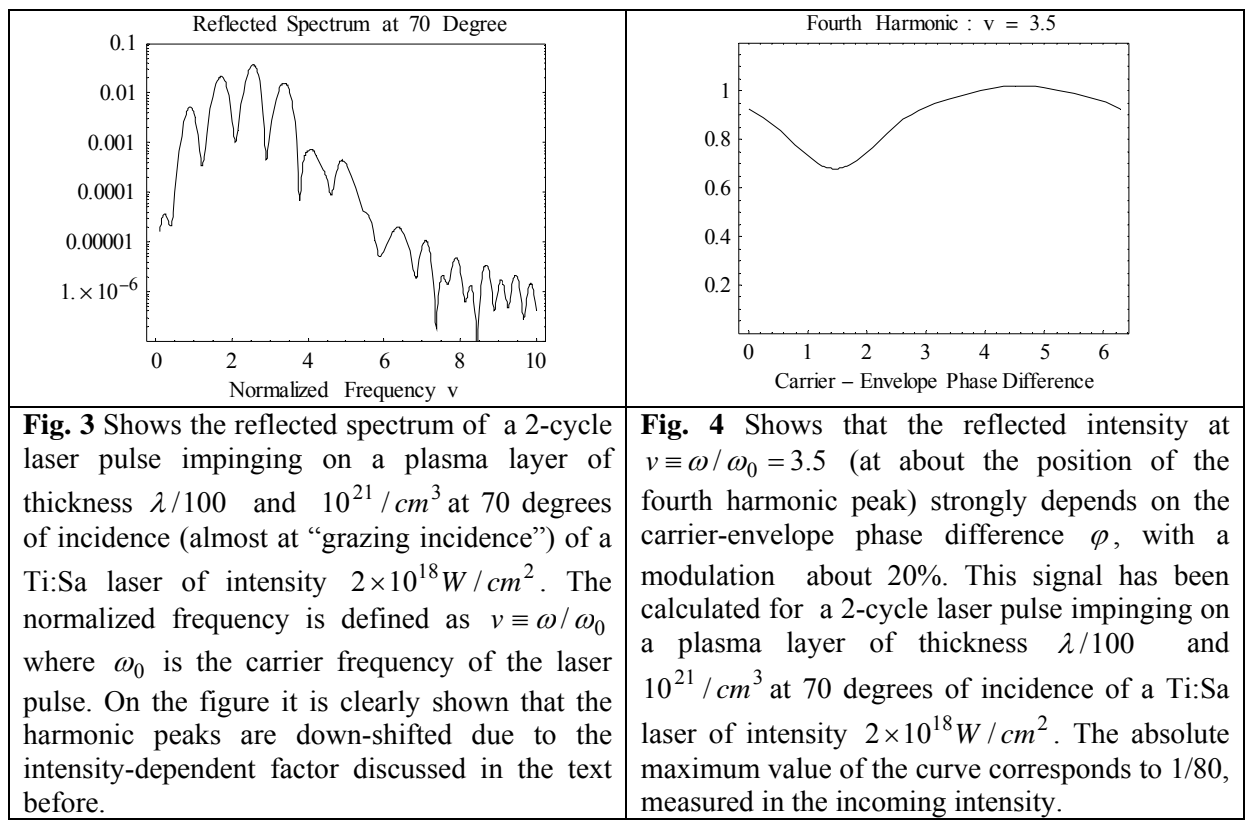

**Fig. 3** Shows the reflected spectrum of a 2-cycle laser pulse impinging on a plasma layer of thickness $\lambda/100$ and $10^{21}/cm^3$ at 70 degrees of incidence (almost at "grazing incidence") of a Ti:Sa laser of intensity $2 \times 10^{18} W/cm^2$. The normalized frequency is defined as $v \equiv \omega/\omega_0$ where $\omega_0$ is the carrier frequency of the laser pulse. On the figure it is clearly shown that the harmonic peaks are down-shifted due to the intensity-dependent factor discussed in the text before.

**Fig. 4** Shows that the reflected intensity at $v \equiv \omega/\omega_0 = 3.5$ (at about the position of the fourth harmonic peak) strongly depends on the carrier-envelope phase difference $\varphi$, with a modulation about 20%. This signal has been calculated for a 2-cycle laser pulse impinging on a plasma layer of thickness $\lambda/100$ and $10^{21}/cm^3$ at 70 degrees of incidence of a Ti:Sa laser of intensity $2 \times 10^{18} W/cm^2$. The absolute maximum value of the curve corresponds to 1/80, measured in the incoming intensity.

On **Fig. 3** we see a typical spectrum of the reflected signal at almos grazing incidence, where the third harmonic dominate. In the next example we illustrate on **Fig. 4** that the maximum value of the 4$^{th}$ harmonic strongly depends on the carrier-envelope phase difference $\varphi$, the modulation (which is defined as $M = (I_{max} - I_{min})/(I_{max} + I_{min})$) is about 20%. We have checked this modulation $M$ for the other spectral ranges, and found that it varies from very small values to 25% for the parameters used in obtaining **Fig. 4**. For the integrated spectrum the value of $M$ is very small, a couple of percents.

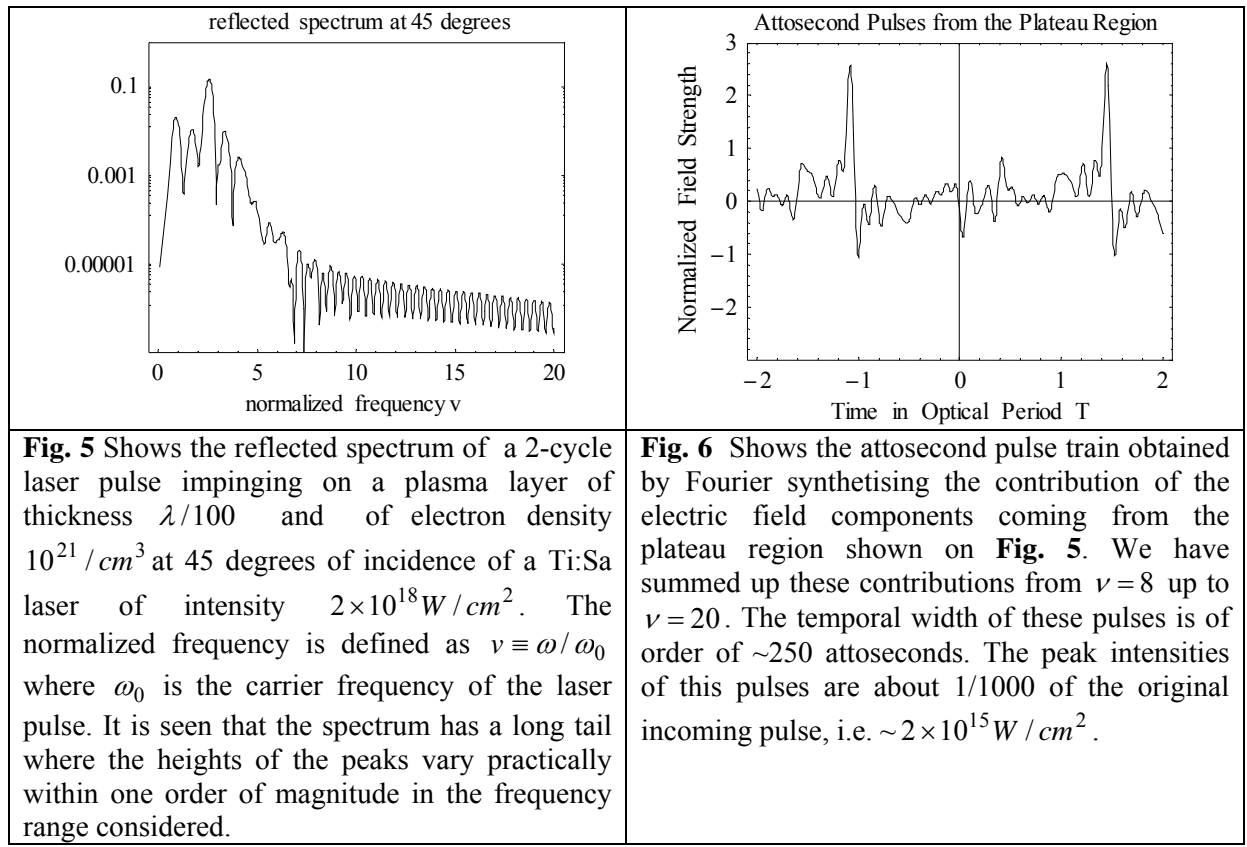

**Fig. 5** Shows the reflected spectrum of a 2-cycle laser pulse impinging on a plasma layer of thickness $\lambda/100$ and of electron density $10^{21}/cm^3$ at 45 degrees of incidence of a Ti:Sa laser of intensity $2\times 10^{18} W/cm^2$. The normalized frequency is defined as $v \equiv \omega/\omega_0$ where $\omega_0$ is the carrier frequency of the laser pulse. It is seen that the spectrum has a long tail where the heights of the peaks vary practically within one order of magnitude in the frequency range considered.

**Fig. 6** Shows the attosecond pulse train obtained by Fourier synthetising the contribution of the electric field components coming from the plateau region shown on **Fig. 5**. We have summed up these contributions from $v = 8$ up to $v = 20$. The temporal width of these pulses is of order of ~250 attoseconds. The peak intensities of this pulses are about 1/1000 of the original incoming pulse, i.e. $\sim 2\times 10^{15} W/cm^2$.

The presence of the plateau shown on **Fig. 5** suggests that by Fourier synthesis of this part of the spectrum results in the appearance of attosecond pulses, as was first proposed by Farkas and Tóth (1992). Note that the spacing of the individual peaks in the plateau region is much smaller than the central frequency, thus this plateau is almost a quasi-continuum. The result of the Fourier syntheses is shown on **Fig. 6**, where we see a part of an attosecond pulse train in the reflected signal. The width of the individual pulses are about on tenth of the optical cycle of the incoming laser pulse, i.e. they have a duration of about 250 attoseconds. Of course, when we synthetise the Fourier components from a wider frequency window, we receive even shorter individual pulses.

## 4. Summary

In the present paper we have described the reflection and transmission of a few-cycle laser pulse on a thin metal layer and a plasma layer represented by a surface current density of free electrons. Our analysis here, as in our earlier study (Varró, 2004), is based completely on classical electrodynamics and mechanics, in the frame of which we solve the system of coupled Maxwell-Lorentz equations of the incoming and scattered radiation and the surface current. In the first case the target can be produced, for instance, by evaporating an aluminum layer on a glass plate. In the second case we can thick of a plasma layer generated from a thin foil in vacuum by a pre-pulse which is followed by the main high-intensity laser pulse.

In Section 2 we presented the basic equations describing our model, and gave the exact analytic solution of the scattering problem in the nonrelativistic regime. Here we have briefly analysed the exact solutions in the frequency domain. In discussing the temporal behaviour of the reflected signal, we found a pulse-decompression and "freezing-in" of the radiation field, yielding to the appearance of a quasi-static wake-field in the scattered signal.

The characteristic time of these wake-fields is inversely proportional to the squared plasma frequency and the thickness of the electron layer. This characteristic time can be much larger than the central period of the incoming laser pulse, hence, it can be much larger then the pulse duration of the original few-cycle pulse. In the numerical example we have taken, the amplitude of the frozen-in wake-field in the reflected signal is of oder of 50000 *V/cm*, which is quite a large value for a quasi-static field. The amplitude and sign of these wake-fields can be varied by changing the CE phase of the incoming pulse. We have pointed out that these wake-fields can perhaps serve as a basis for the construction of a robust linear carrier-envelope phase difference meter.

In Section 3 we derived the relativistic equation of motion for the surface current density of the electrons which are considered the active charges in a thin plasma layer. It turned out that the complete solution of the scattering problem can be reduced to the solution of a first order ordinary (nonlinear) inhomogeneous differential equation. An approximate analytic solution to this equation has been given, which is valid for moderately relativistic incoming laser intensities. On the basis of these solutions the spectrum of the reflected radiation field containing higher-harmonics have been calculated, and we have seen some illustrative numerical examples summarized on the figures 3-6. A main characteristics of these spectra is the intensity-dependent frequency down-shift of the harmonic peaks. This effect is an analogon of the well-known intensity dependent frequency shift in the nonlinear Thomson scattering on a single electron. For longer pulse we have found a sharp cut-off in the harmonic spectra, whose position can well be estimated from the theory of synchrotron radiation. It has also been shown that the spectra considerably depend on the CE phase of the incoming few-cycle laser pulse. In certain regions of the reflected spectra we have found about twenty percent of modulation (visibility) by varying the carrier envelope phase difference. In general, the high-harmonic spectra have a plateau, and the Fourier components of this quasi-continuum are locked in phase. As a consequence, by synthesing this part of the spectrum we have obtained attosecond pulses in the reflected signal.The widths of the individual pulses, coming from a frequency window taken in our numerical example, were about on tenth of the optical cycle.

**Acknowledgements.** This work has been supported by the Hungarian National Scientific Research Foundation, OTKA, grant no. T048324.

**References**
Apolonskiy, A., Dombi, P., Paulus, G. G., Kakehata, M., Holtzwarth, R., Udem, Th., Lemell, Ch., Torizuka, K., Burgdörfer, J., Hänsch, T. W. & Krausz, F. (2004). Observation of light-phase-sensitive photoemission from a metal. *Phys. Rev. Lett.* **92**, 073902 (1-4).

Baltuška, A., Udem, Th., Uiberacker, M., Hentschel, M., Goulleimakis, E., Gohle, Ch., Holtzwarth, R., Yakovlev, V. S., Scrinzi, A., Hänsch, T. W. & Krausz, F. (2003). Attosecond control of electronic processes by intense light fields. *Nature* **421**, 611-615.

Brabec, T. & Krausz, F. (2000). Intense few-cycle laser fields: Frontiers of nonliear optics. *Rev. Mod. Phys.* **72** , 545-591.

Chelkowski, S. & Bandrauk, A. (2002). Phase-dependent asymmetries in strong-field photoionization by few-cycle laser puses, *Phys. Rev.* A **65** , 061802 (1-4).

Dombi, P., Apolonskiy, A., Lemell, Ch., Paulus, G. G., Kakehata, M., Holtzwarth, R., Udem, Th., Torizuka, K., Burgdörfer, J., Hänsch, T. W., & Krausz, F. (2004). Direct measurement and analysis of the carrier-envelope phase in light pulses approaching the single-cycle regime. *New J. Phys.* **6**, 39-48.

Sándor Varró: Linear and Nonlinear Absolute Phase Effects                    17

Drescher, M., Hentschel, M., Kienberger, R., Uiberacker, M., Yakovlev, V. S., Scrinzi, A., Westerwalbesloh, Th., Kleineberg, U., Heinzmann, U. & Krausz, F. (2002). Time-resolved atomic inner-shell spectroscopy. *Nature* **419**, 803-807.

Farkas, Gy. & Tóth, Cs. (1992). Proposal for attosecond light pulse generation using laser-induced multiple-harmonic conversion processes in rare gases. *Phys. Lett. A* **168**, 447-450.

Fearn, H. & Lamb, Jr., W. E. (1991). Correction to the golden rule. *Phys. Rev.* A **43**, 2124-2128.

Fortier, T. M., Roos, P. A., Jones, D. J. & Cundiff, S. T. (2004). Carrier-envelope phase-controlled quantum interference of injected photocurrents in semiconductors. *Phys. Rev. Lett.* **92**, 147403 (1-4).

Gradshteyn, I. S. & Ryzhik, I. M. (1980) *Table of Integrals, Series and Products.* p. 307. New York: Academic Press.

Hentschel, M., Kienberger, R., Spielmann, Ch., Reider, G. A., Milosevic, N., Brabec, T., Corkum, P., Heinzmann, U., Drescher, M. & Krausz, F. (2001). Attosecond metrology. *Nature* **414**, 509-513.

Hidding, A., Amthor, K.-A., Liesfeld, B., Schwoerer, H., Karsch, S., Geissler, M., Veisz, L., Schmid, K., Gallacher, J. G., Jamison, S. P., Jaroszynski, D., Pretzler, G. & Sauerbrey, R. Generation of quasimonoenergetic electron bunches with 80-fs laser pulses. (2006). *Phys. Rev. Lett.* **96**, 105004 (1-4).

Jackson, J. D. (1962). *Classical Electrodynamics*. Section 14.6. New York: John Wiley & Sons, Inc.

Kiselev, S., Pukhov, A. & Kostyukov, I. (2004). X-ray generation on strongly nonlinear plasma waves. *Phys. Rev. Lett.* **93**, 135004 (1-4).

Lemell, Ch., Tong, X.-M., Krausz, F. & Burgdörfer, J. (2003). Electron emission from metal surfaces by ultrashort pulses: Determination of the carrier-envelope phase. *Phys. Rev. Lett.* **90**, 076403 (1-4).

Lichters, R., Meyer-ter-Vehn, J. & Pukhov, A. (1996). Short-pulse laser harmonics from oscillating plasma surfaces driven at relativistic intensity. *Phys. Plasmas* **3**, 3425-3437.

Milošević, D. B., Paulus, G. G. & Becker, W. (2002). Phase-dependent effects of a few-cycle laser pulse. *Phys. Rev. Lett.* **89**, 15300 (1-4).

Milošević, D. B., Paulus, G. G. & Becker, W. (2003). High-order above-threshold ionization with a few-cycle laser pulse: a meter of the absolute phase. *Optics Express* **11**, 1418-1429.

Nakajima, T. & Watanabe, Sh. (2006). Effects of the carrier-envelope phase in the multiphoton ionization regime. *Phys. Rev. Lett.* **96**, 213001 (1-4).

Paulus, G. G., Grasbon, F., Walther, H.,Villoresi, P., Nisoli, N., Stagira, S., Priori, S., & De Silvestri, S. (2001). Absolute-phase phenomena in photoionization with few-cycle laser pulses. *Nature* **414**, 182-184.

Pukhov, A. & J. Meyer-ter-Vehn, J. (2002). Laser wake field accelerator: the highly non-linear broken-wave regime. *Appl. Phys. B* **74**, 355-361.

Quèrè, F., Thaury, C., Monot, P., Dobosz, P., Martin, Ph., Geindre, J.-P. & Audebert, P. Coherent wake emission of high-order harmonics from overdense plasmas. (2006). *Phys. Rev. Lett.* **96**, 125004 (1-4).

Ristow, T. (2004). *Anregung von Atomen durch Laserpulse mit wenigen Zyklen.* Diplom work. Aachen (Germany): Rheinisch-Westfälischen Technischen Hochschule.

Sommerfeld, A. (1915). Über das Spektrum der Röntgenstrahlung. *Ann. der Physik*, **46**, 721-747.

Varró, S. (2004). Scattering of a few-cycle laser pulse on a thin metal layer: the effect of the carrier-envelope phase difference. *Laser Phys. Lett.* **1**, 42-45.

Witte, S., Zinkstok, R. T., Hogerworst, W. & Eikema, K. S. E. (2004). Control and precise measurement of carrier-envelope phase dynamics. *Appl. Phys. B* **78**, 5-12.